\documentclass[conference,compsoc]{IEEEtran}
\usepackage{cite}
\usepackage{amsfonts}
\usepackage{amsmath}
\usepackage{amssymb}
\usepackage{subfig}
\usepackage{color}
\usepackage{graphicx}
\usepackage[hyphens]{url}
\usepackage{mdwlist}
\usepackage{xspace}
\usepackage{cite}

\usepackage{soul}          %
\usepackage[draft,inline,nomargin]{fixme}

\graphicspath{{images/}{charts/}}
\DeclareGraphicsExtensions{.pdf}
\newcommand{\dosn}{DECENT\xspace}

\newcommand{\squishlist}{
 \begin{list}{$\bullet$}
	{ \setlength{\itemsep}{0pt}
    \setlength{\parsep}{3pt}
    \setlength{\topsep}{3pt}
    \setlength{\partopsep}{0pt}
    \setlength{\leftmargin}{1.0em}
    \setlength{\labelwidth}{1em}
    \setlength{\labelsep}{0.5em}} }
\newcommand{\squishend}{\end{list}}

\begin{document}

\title{\dosn: A Decentralized Architecture for Enforcing Privacy in Online Social Networks}

\author{Sonia Jahid$^{\textrm{\small 1}}$ \hspace{1em}
Shirin Nilizadeh$^{\textrm{\small 2}}$ \hspace{1em}
Prateek Mittal$^{\textrm{\small 1}}$ \hspace{1em}
Nikita Borisov$^{\textrm{\small 1}}$ \hspace{1em}
Apu Kapadia$^{\textrm{\small 2}}$ \\[1ex]
$^{\textrm{\small 1}}$\parbox[t]{2.5in}{\small University of Illinois at Urbana-Champaign\\
\{sjahid2,mittal2,nikita\}@illinois.edu}
\hspace{1em}
$^{\textrm{\small 2}}$\parbox[t]{1.3in}{\small Indiana University  \\
\{shirnili,kapadia\}@indiana.edu}
}

\maketitle

\begin{abstract}

A multitude of privacy breaches, both accidental and malicious, have prompted users to distrust centralized providers of online social networks (OSNs) and investigate decentralized solutions. We examine the design of a fully decentralized (peer-to-peer) OSN, with a special focus on privacy and security. In particular, we wish to protect the confidentiality, integrity, and availability of user content and the privacy of user relationships. We propose \dosn, an architecture for OSNs that uses a distributed hash table to store user data, and features cryptographic protections for confidentiality and integrity, as well as support for flexible attribute policies and fast revocation. \dosn ensures that neither data nor social relationships are visible to unauthorized users and provides availability through replication and authentication of updates. We evaluate \dosn through simulation and experiments on the PlanetLab network and show that \dosn is able to replicate the main functionality of current centralized OSNs with manageable overhead. 

\end{abstract}
\section{Introduction}
\label{sec:introduction}

Online social networks (OSNs) such as Facebook and Google+ have revolutionized 
the way people interact and are being used by hundreds of millions of users 
across the world. However, these designs suffer from a key problem: 
the lack of user privacy. First, users are not in control of their private 
data---the social network provider has full access to the user's data, and, 
in fact, several providers have been caught selling user data~\cite{osn-selling-data}. 
Second, these networks do not enable a user to set fine-grained policies for 
access control. For example, users can control access to a status messages in Facebook by using lists (and likewise by using ``circles'' in Google+), no policy can be defined for comments and actions such as Likes or +1's. %
 The problem is further exacerbated by the network provider's 
constantly changing and oblique privacy policies~\cite{eff-facebook-timeline}. Recent accidental 
and malicious privacy breaches have further motivated designs that enable social networking with privacy preservation~\cite{flickr, gross, app:facebook}.  

Unfortunately, alternative proposals to improve user privacy in OSNs do not provide an 
adequate level of privacy. Proposals for centralized designs~\cite{Singh09} either trust 
the OSN provider with user data and/or allow the provider to perform effective traffic 
analysis attacks to undermine user privacy by, for example, learning a user's social contacts.
Users consider their social contacts to be sensitive information, evidenced by the public 
outcry when Google Buzz made this information public, and was forced to change its 
default settings~\cite{google-buzz}. %
While alternative decentralized designs~\cite{buchegger09,diaspora,lotusnet,Cutillo09} do 
not rely on a single trusted or untrusted entity, such designs do not focus on data management, and data are usually stored by the owners themselves or their trusted contacts. %

We propose \dosn, an architecture for enforcing access control in a decentralized OSN. 
Our focus is on providing data confidentiality, integrity, and availability in the presence of 
malicious nodes in a distributed setting.
 Our architecture is also able to protect the privacy of user relationships. \dosn is based around 
a flexible object-oriented design (OOD) that supports the main functionality of OSNs and captures 
the complex multi-principal interactions that are common in social networks. The confidentiality 
and integrity of data are protected by a cryptographic mechanism so that they can be stored in 
untrusted nodes in a distributed hash table (DHT).  The standard DHT mechanisms are extended to 
ensure availability despite malicious attempts to erase or overwrite stored data.

Our contribution in this paper is twofold:

\textbf{1) Design:} We propose a decentralized OSN architecture that: i) provides flexibility in data management 
through OOD; ii) uses an appropriate and advanced cryptographic scheme that supports efficient access revocation and fine-grained 
policies on each piece of data; and iii) combines confidentiality, 
integrity, and availability by using the functionalities of a DHT---all the existing 
designs focus on one or two, but not all of these aspects. The novelty of our architecture 
lies in integration of existing primitives that are tailored to enhance the security and 
privacy of OSNs.

\textbf{2) Prototype:} We develop a prototype of \dosn---the wall and newsfeed functionalities,
to be specific---and evaluate its performance through simulation and experiments on PlanetLab. 
We evaluate \dosn using a FreePastry simulator and a Kademlia implementation on PlanetLab~\cite{Rowstron01pastry:scalable, Maymounkov02}. %
Our preliminary analysis shows 
that the overhead of \dosn is moderate, and that our architecture for privacy-preserving decentralized OSNs is feasible. We are currently investigating techniques to further improve performance 
such as cryptographic optimizations~\cite{green:usenix11}.
Our architecture thus demonstrates that existing security primitives with well understood 
properties can be leveraged to provide a compelling privacy-preserving alternative to centralized OSNs. 

\section{Requirements and Properties}\label{sec:requirement}

\subsection{Functional Model}

Much of the functionality of OSNs can be described as users posting content and their social contacts viewing, commenting on, and annotating  such content.  To provide a flexible, general model of these operations, we define a \emph{container object} that has two components: the main content and a list of comments/annotations, represented as references to other container objects.  The main content can take on many types, such as a status update, a shared link, a photo or video, or a collection of container objects (e.g., a photo album).  For our purposes, the content type is not important; the key difference is that the \emph{access permissions} on the comments can be more restrictive than the content to enforce the policy of each object individually.

A user's profile is a root object, which contains references to other objects, such as contact information, a wall, photo albums, etc. 
Similarly, other objects may consist of some content and references to other objects. For example, a wall may have references to status messages and posts where a status/post object may contain references to comment objects in addition to the status data. Thus, each user's content is organized in a hierarchical fashion (although we do not enforce a tree structure---a single object may be referenced by multiple ``parent'' objects). 
With this degree of granularity in our object design, access permissions can be assigned specifically for each object and then referenced by other containers. 

\subsection{Security Requirements}

Within this model, we can outline a number of security and privacy requirements:

\textbf{Confidentiality}: Preserving the confidentiality of user content is a key requirement for a decentralized OSN. Content should be accessible to only those who are explicitly authorized by the content owner. Furthermore, nodes hosting such data may themselves not be authorized to read the data. 

\textbf{Integrity}: We must also ensure the integrity of the data so that OSN users can be certain that content posted by their friends is authentic. This property is important in a peer-to-peer network since storage nodes are untrusted and may try to perform unauthorized updates to the stored data.

\textbf{Availability}: User content should remain available until it is explicitly deleted by its owner, even if the owner is offline, and despite potential malicious attempts to destroy the data.  Readers should also be able to retrieve the most recent version of a content object rather than past ones.

\textbf{Discretionary Control}: Policies controlling who may view, modify, or comment on content are 
defined by its owner and cannot be changed without the owner's authorization.

\textbf{Flexible Policies}: Users should be able to define fine-grained access policies that can be phrased in 
terms of %
a conjunction or disjunction of attributes given to social contacts~\cite{Baden09} (e.g.,``(friend AND co-worker) OR family''), as well as relationship 
degree (e.g., ``friends of friends'').

\textbf{Relationship Privacy}: Relationships between users %
should remain hidden from third parties that may have no relationship with the 
object owner and are therefore untrusted, such as storage nodes.

\subsection{Threat Model}

We assume that the participants in the decentralized OSN may be malicious (or compromised) 
and therefore should not be trusted with the confidentiality and integrity of data and relationship information. Moreover, the malicious entities are considered to be 
Byzantine, and can launch both active and passive attacks. Distributed systems are 
vulnerable to the problem of Sybil attacks, where a single entity can obtain multiple 
identities in the system and violate security properties. We assume the existence of 
mechanisms to defend against the Sybil attack~\cite{castro,whanau}. 
We consider that up to 25\% of the nodes in the system can be malicious, since, beyond that, 
existing mechanisms~\cite{castro} are not able to securely route in distributed hash tables, which is a 
necessary prerequisite to provide both integrity and availability guarantees. In this paper we focus on the cryptographic mechanisms to protect the security of stored data, rather than routing-based attacks, which can be addressed by existing mechanisms, and which we will therefore not discuss further.

\section{System Architecture}\label{sec:arch}

\dosn is a decentralized OSN, 
which employs a DHT to store and retrieve data objects created by their owners. 
 Each object is encrypted to provide confidentiality. The primary advantage of our architecture is its modularity, i.e., the data objects, the cryptographic mechanisms, and the DHT are three separate components, interacting with each other through well-defined interfaces. The modular design provides us with the capability of using any type of DHT or cryptographic scheme for the prototype implementation. 

\subsection{Access Policies}

Each object has three access policies associated with it. The policies are either attribute-based (AB), 
 identity-based (IB), or a combination of both types.  AB policies take the standard form of policies described through various attributes, for example, \emph{friend}, \emph{family}, \emph{coworker}.  AB policies can represent formulas over attributes, using operators such as $\wedge$, $\vee$, and $k$-of-$n$. Examples of AB policy are: (\emph{friend} $\wedge$ \emph{coworker}) $\vee$ \emph{family}, and 2~\textit{of}~$\{\mathit{friend}, \mathit{family}, \mathit{coworker}\}$ (background on Attribute-based Encryption (ABE) is provided in section~\ref{crypto}).

\squishlist
\item Read policy (\emph{R-Policy}) describes who may read the contents of the object. It is an AB policy that describes the attribute combination required for a user to decrypt an object's data. 

\item Write policy (\emph{W-Policy}) describes who may modify the contents of the object or delete the object. It is an IB policy, which generally is set to the owner of the object.

\item Append policy (\emph{A-Policy}) describes who may add a comment/annotation to the object. It is also an AB policy.
\squishend            

These policies are defined by the owner at the time of object creation and are stored in the object metadata.
The read policy is enforced through the use of cryptography.  The write and append policies are enforced through a combination of cryptography and specialized DHT functionality. A reader can cryptographically verify the integrity of the object and be sure that the content and comments have been posted by parties authorized by the write and append policies, respectively. Additionally, the DHT storage nodes require authorization before each write operation to prevent malicious deletions and vandalism. 
The authorization does not reveal a user's identity, hence the storage node is not aware of the identities of users storing or retrieving data from it, and therefore a user's social graph is hidden from the storage nodes.  DHT nodes also implement a special append operation that adds a new annotation to the object while leaving existing content unmodified.
When objects are being stored at malicious nodes, confidentiality is still preserved due to cryptography. However, the malicious
nodes can impact the integrity and availability guarantees by deviating from the protocol, e.g., by deleting objects and/or
returning previous versions of objects.  Malicious nodes in the DHT can be tolerated using replication. 

\subsection{Cryptographic Protection}
\label{crypto}

Objects are stored on untrusted DHT nodes, thus necessitating the use of cryptography to protect their confidentiality and integrity.  For confidentiality, Baden et al.~\cite{Baden09} observed that ABE~\cite{cpabe} is a good fit for OSNs because it allows users to specify access policies in terms of groups of contacts, such as friends, family, coworkers, etc.\ We adopt several modifications to the base ABE to better satisfy our security requirements.
                       
\textbf{Confidentiality:} Attribute-based encryption is a public-key encryption scheme where each encrypted item is associated with a policy.  There are many decryption keys, and each one is associated with a set of attributes.  A key can decrypt an encrypted item if its set of attributes satisfies the item's policy.
A key authority (KA) maintains the master secret key (MSK) and can generate decryption keys with an arbitrary set of attributes.  In the OSN context, each user becomes a key authority, issuing different encryption keys to social contacts based on their attributes.  The social contact will know which attributes identifiers they possess, but not their semantic meaning.

Like most public-key schemes, ABE is usually used in hybrid encryption mode, wherein the message is encrypted with a randomly chosen symmetric encryption key, which is in turn encrypted with ABE.  We follow this approach with a modification that the ABEncrypted symmetric key is in fact part of the object reference and not included in the object itself.  The main motivation for this choice is that the version of ABE we are using lacks %
policy privacy and this approach keeps %
the policy hidden from untrusted storage nodes . %

An additional consequence of this approach is that when several references for an object exist, the object may have different read policies associated with it.  For example, if Bob posts a comment on a status update on Alice's wall, he may wish to add a reference to his comment (or even the status update) to his own wall so that it can be seen by his contacts.\footnote{Note that this may contravene Alice's privacy wishes, but no architectural protection short of DRM can prevent Bob from such re-sharing. A possible mitigation strategy is to issue a warning 
to Bob about the possible privacy breach, as is currently implemented in Google+.}  In this case, Bob's comment may be visible to some subset of Alice's contacts when reached through her status update, and some subset of Bob's contacts when reached through his wall.

A second extension of ABE is the support for immediate revocation by the use of the EASiER scheme~\cite{easier}.  From time to time, a user may wish to revoke one or more attributes from a social contact.  An ideal revocation scheme would ensure that the revoked contact can no longer access any data that requires the revoked attribute(s), including existing data.  To support this, EASiER makes use of a proxy in every decryption.  In brief, decryption keys in EASiER are blinded in a way specific to each user's identity; to decrypt a data item, a user with the appropriate key must contact the proxy to transform the ciphertext so that it is compatible with the blinded key.  When Alice wishes to revoke an attribute, she updates the proxy key in such a way that this transformation is no longer possible for the revoked users.  The proxy itself is minimally trusted---it can neither decrypt the data nor restore access for a previously revoked user even if it is compromised.

We propose two extensions to the base EASiER scheme.  First, we use threshold secret sharing to split the proxy functionality among several randomly selected nodes; since we assume that the majority of nodes are not actively malicious, it will ensure the security of the proxy.  Second, we extend EASiER to support attribute delegation: if Alice issues a key with a set of attributes to Bob, Bob can delegate a subset of these attributes to Carol. This way, Alice can define a ``friend-of-a-friend'' attribute and ask all her contacts to delegate it to all of their contacts.  Note that Carol will have to use both Alice's and Bob's proxies for decryption, so if either Alice revokes Bob's access or Bob revokes Carol's, the decryption will fail.  These extensions will be described in an upcoming technical report.    
Note that the write policy public key must be part of the object reference, rather than the object itself, to ensure its authenticity. The append policy, on the other hand, is included as part of the object metadata, since it is authenticated by the write-policy signature.  An object reference, therefore, consists of:

\[ \mathit{objRef} \stackrel{\mathrm{def}}{=} (\mathit{objID}, \mathrm{ABE}(K, P), \mathit{SPK}) \]

\noindent where \emph{objID} is a random object identifier, used to locate it in the DHT, $K$ is the symmetric key used to encrypt the referenced object, $P$ is the attribute-based read policy, and \emph{SPK} is the write-policy signature public key.  To optimize storage requirements, the \emph{SPK} can be omitted, implying that the write policy of the referenced object is the same as of the containing one; likewise, ABE($K,P$) can be replaced by an unencrypted $K$ if the read policy of the referenced object matches the container.  
We note that these references are similar in spirit to capabilities used in  Tahoe-LAFS~\cite{tahoe}.

\subsection{Distributed Hash Table}

Participants in the OSN are organized into a distributed hash table (DHT), such as Pastry~\cite{Rowstron01pastry:scalable} or Kademlia~\cite{Maymounkov02}.  The DHT creates a scalable key-value store with an efficient lookup mechanism to locate nodes that store a given object.  DHT lookups can be made secure against malicious attacks~\cite{castro}, ensuring that a lookup will find the correct copy of an object if it exists.  (It may additionally find incorrect copies provided by malicious nodes; cryptographic integrity protection described above can be used to identify the correct one.)

Objects in \dosn are stored in the DHT using the \textit{objID} as the key. 
To ensure availability despite node churn and malicious attacks, several replicas of an object are maintained.  DHTs typically use the neighbor set of the node responsible for the object key to maintain replicas; the number of replicas needs to be tuned based on the churn patterns of the network (malicious nodes can also be modeled as churn in this case), which we will study in our future evaluation. To guarantee freshness, each object has a version number as part of its metadata. The version number is authenticated by the write-policy signature, thus a user can query all of the replicas and use the freshest object returned.  

Malicious users may try to modify or delete an existing object.  Note that the write policy prevents them from creating modifications that will be accepted by the readers, as they cannot produce a correct signature.  The storage node, however, does not know the \emph{SPK}, as it is part of the object reference, and thus cannot distinguish a legitimate update from a malicious one that overwrites and destroys user data.

To address this issue we add an unencrypted metadata field to an object containing a public key that is used to authenticate write requests (\emph{write authentication public key}, or \emph{WAPK}).  Any write or delete request for an existing object must be 
signed by the corresponding secret key; otherwise, the storage node will refuse the request.  Thus, as long as there is always at least one honest (but curious) replica for an object, it will persist despite any malicious attacks.  This public key \emph{should not} be a user's permanent public key, as otherwise a storage node could use it to link an object to its owner.
Instead, a separate \emph{WAPK} is generated for each object, ensuring unlinkability of objects.  A copy of the corresponding secret key (\emph{WASK}) is stored in the object, encrypted with the owner's secret key.  We use Digital Signature Algorithm (DSA) for write access signatures, which can be generated very quickly and should not create a performance bottleneck.

In addition to the standard \emph{get} (read) and \emph{put} (write) requests, the \dosn DHT supports an \emph{append} request, which is used to add a comment on an existing object.  Note that again, the storage nodes do not know the append-policy signature public key, as it is part of the encrypted object metadata.  Unauthorized appends, however, can simply be discarded by the readers of the object and do not affect availability.

\subsection{Example}

\textbf{Join:} To join \dosn , Alice sets up her profile, wall, and keys. She generates her ABE public and master secret keys and signature key pair. Alice creates an object with her profile information, encrypts it with a symmetric key, and signs it with her write-policy signature key $SPK_{Alice}$. She generates a random ID, saves her profile in the DHT using this ID, ABEncrypts the symmetric key with profile R-policy, and creates a reference to the profile object. Similarly, she sets up her wall. Alice can be reached through the reference to a root object, which contains the references to her profile and wall. The root object acts as another regular object and is stored in the DHT. The root object can be thought of a user's landing page on Facebook. 

\textbf{Establish Contacts:} To establish the relationship \emph{friend, co-worker} with Bob, Alice generates an ABE secret key for Bob with the attributes \emph{friend, co-worker}. Relationships are asymmetric,  so Bob may establish just \emph{acquaintance} relationship with Alice by issuing an ABE secret key for this attribute. Keys are exchanged out of band. Alice and Bob also exchange their root object references.

\begin{figure}[htbp]
	\centering
		\includegraphics[width=0.8\columnwidth]{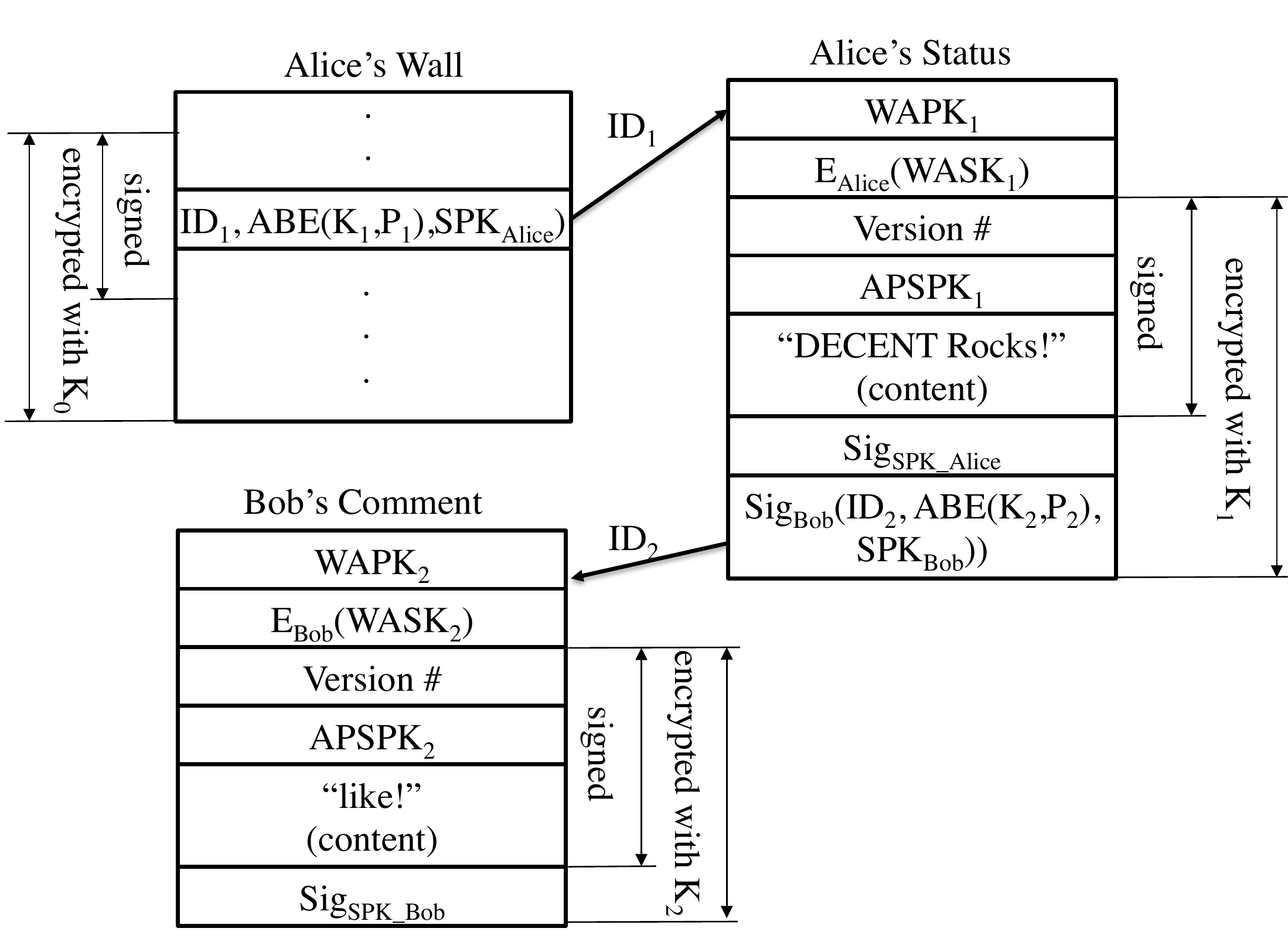}
	\caption{Example objects}
	\label{fig:obj}
\end{figure}

\textbf{Post and Comment: } Figure~\ref{fig:obj} shows an example object structure.  When Alice wants to post a status update to her wall, she creates the status update object, complete with version number, contents, and append policy ($\mathit{APSPK_1}$), and generates a signature 
over these three values using the secret key corresponding to a write-policy signature key ($SPK_{Alice}$).  She also generates a random write-authentication public and private key ($\mathit{WAPK_1}$ and $\mathit{WASK_1}$) and stores them in the object, encrypting $\mathit{WASK_1}$ to herself.  She then picks a random symmetric encryption key $K_1$ and encrypts the object (except for  $\mathit{WAPK_1}$ and $\mathit{WASK_1}$); she also chooses a random id $\mathit{ID}_1$ and uses this to insert the object into the DHT.  Finally, she creates a reference to the status update, including $\mathit{ID}_1, K_1$ and her write-policy public key ($SPK_{Alice}$) and adds it to her wall. $K_1$ is encrypted with an AB policy $P_1$ (R-policy discussed before), which governs who can read the status update. Note that Alice's wall is also encrypted in a similar way with another random key $\mathit{K_0}$.

When Bob wants to read Alice's update, he finds the reference on Alice's wall and decrypts $K_1$ with his attribute-based secret key that he got from Alice, assuming that his attributes satisfy the policy $P_1$.  (This decryption also involves Alice's proxy in support of EASiER revocation.)  He then retrieves the object from the DHT with the key $\mathit{ID_1}$ and decrypts the encrypted fields using $K_1$.  Finally, he verifies Alice's signature to ensure the authenticity of the update. Note that Bob has to satisfy the R-Policy associated with the wall object itself to get access to the post or status references in it.

If Bob further wants to comment, then he first creates a comment object following a process similar to Alice's creation of her update.  He then uses the \emph{append} operation to insert a reference to the new object into Alice's update.  The reference is signed by Bob's key that satisfies the A-Policy of the status. Readers will verify that the signature matches $\mathit{APSPK_1}$ in Alice's status A-policy and discard it otherwise. The encryption key $K_2$ is further ABEncrypted using Bob's policy $P_2$; thus, only users who satisfy both $P_1$ and $P_2$ will be able to read the comment.

\section{Implementation and Evaluation}\label{sec:eval}

We have implemented a preliminary prototype of \dosn, which provides functionality similar to the Facebook wall. It also provides a basic newsfeed option, summarizing status updates from a person's contact/friend list. 
We use four different types of cryptographic schemes in \dosn: EASiER for ABE, AES for symmetric encryption, DSA for signatures, and RSA to encrypt the write policy signature key. We use a combination of EASiER and DSA to realize ABS. The key sizes are chosen as recommended by NIST~\cite{nist} for maximum security. We use FreePastry with Euclidean network topology for simulation, and Kademlia~\cite{aiello08, Likir} %
 for the experiments on PlanetLab as the underlying DHT. Our proxy was run on a standard server for simulation and PlanetLab experiments.

\subsection{Simulation}
\begin{figure*}[htbp]
\centering
\subfloat[View My Wall] { \includegraphics[width=0.3\textwidth]{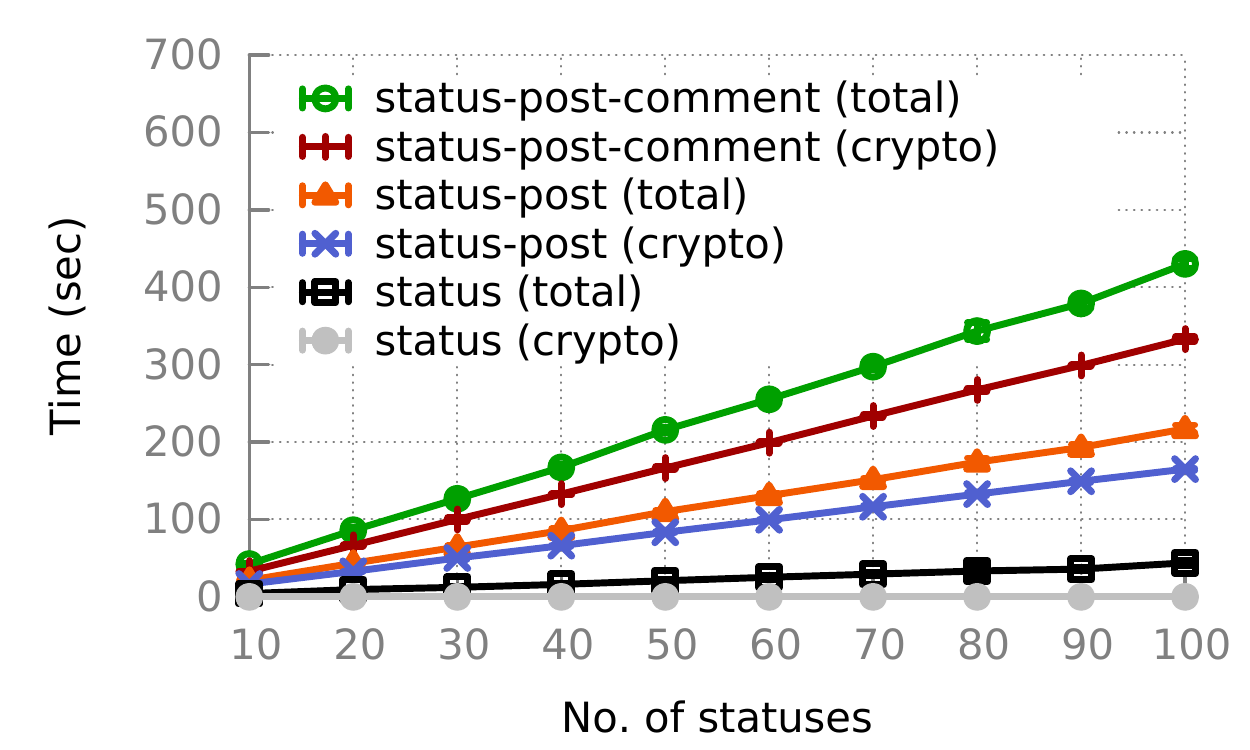} \label{chart:view-my-wall} }
\subfloat[View Others' Wall]{ \includegraphics[width=0.3\textwidth]{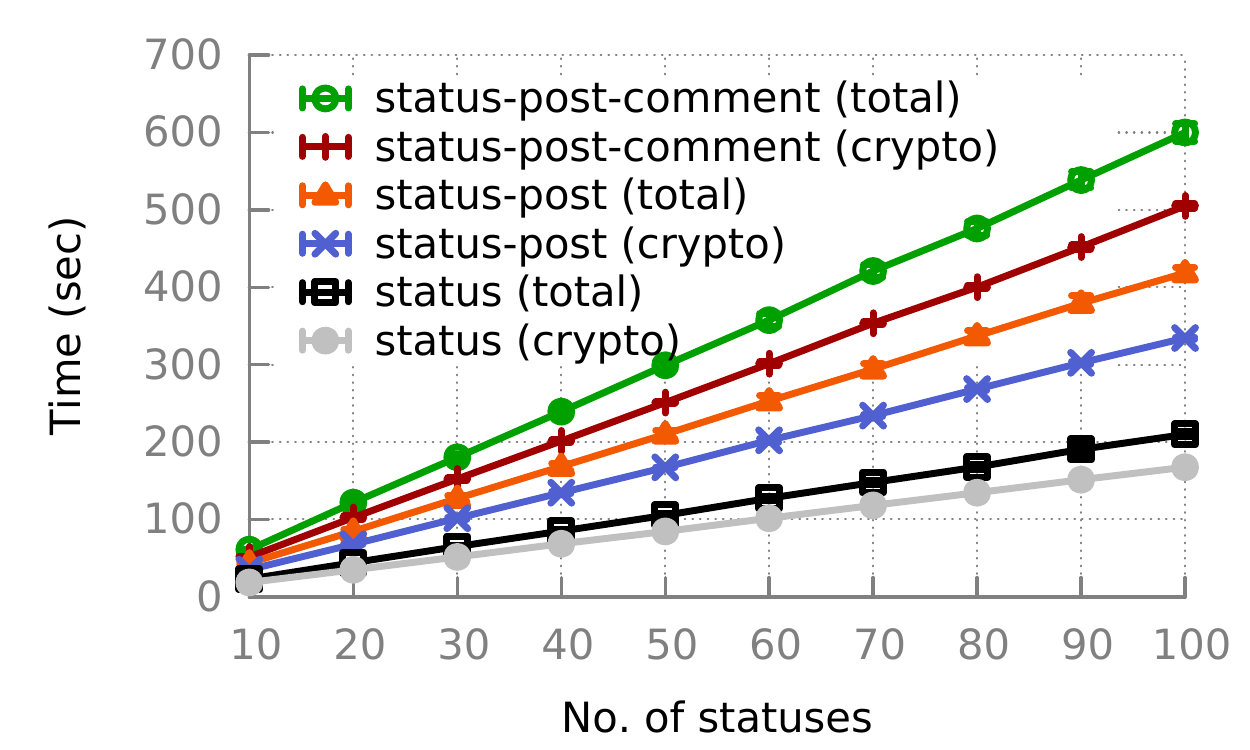} \label{chart:view-others-wall} }
\subfloat[View Newsfeed] {\includegraphics[width=0.3\textwidth]{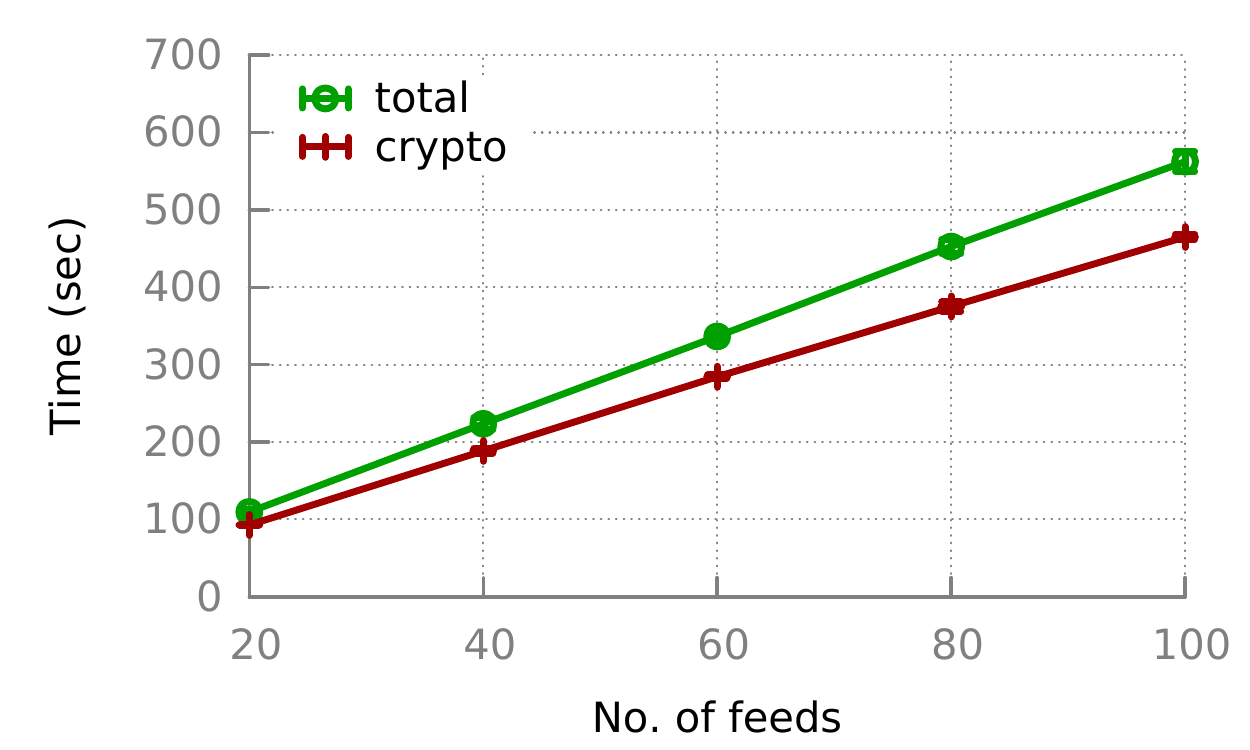} \label{chart:view-news-feed} }
\vspace{-2mm}
\caption{Simulation Results for 10,000 nodes: The average time to view others wall with 10 and 20 statuses/posts/comments is about 
60~seconds and 120~seconds respectively. The average time to view a newsfeed with 20 peers is 109~seconds.}
\label{chart:sim-pastry}
\vspace{-5mm}
\end{figure*}

We perform experiments to measure the performance of viewing a user's newsfeed and wall with varying numbers of status messages, posts, and comments. 
The simulation was run on a peer-to-peer network of $10\,000$ nodes. Figure~\ref{chart:sim-pastry} shows our simulation results.

{\em View Wall}: To view a wall, a user uses the wall reference to fetch the wall object, ABDecrypts the wall reference to get the AES key to decrypt the wall object, and gets the wall metadata and the references to statuses and posts. Each reference is used to fetch the corresponding status or post. The user ABDecrypts the reference to view the content of the object. 

We perform tests to view wall with: 1) only statuses, 2) statuses and same number of posts from friends, and 3) statuses, posts, and one comment on each status from friends. Figures~\ref{chart:view-my-wall} and \ref{chart:view-others-wall} show results for viewing one's own wall and friends' walls respectively. A data point such as $x$ statuses means that when a user views a wall that contains statuses, posts, and comments, she views $x$ of each, i.e., $3x$ objects.

We allow users to cache the AES encryption keys for the objects they create and thus  avoid ABDecryption of the references to their own objects. Therefore, viewing one's own wall with only statuses is much faster than viewing a friend's wall with only statuses. Viewing one's own wall with statuses and posts involves ABDecryption for the posts from friends and only AESDecryption for the statuses. The same applies to comments from friends. When a user has not posted anything herself to her friends' walls, then viewing friends' walls involves ABDecryption for each item on the wall, and so represents the worst case scenario.
The current view time (e.g., $90$\,s to view a user's own wall with $20$ statuses, $20$ posts, and $20$ comments) may appear large; however, content can be displayed progressively, and thus older messages can be fetched while the user is reading the most recent messages, which are loaded within the first few seconds.  We are also currently working on cryptographic optimizations to speed up these operations.

{\em View Newsfeed}: We test our prototype to evaluate the basic newsfeed functionality. This  approach fetches the latest status from each of a user's friends. Figure~\ref{chart:view-news-feed} shows the results. An example newsfeed with $40$ feeds takes around $215$\,s to construct and view. The results will be improved with parallel lookups and decryption. However, in current OSNs a user's newsfeed generally shows 20--30 posts at a time. Some techniques, such as showing feeds in blocks and pre-fetching the latest updates from friends while the user is offline, will improve the performance. We will investigate these techniques in our future evaluation.

{\em Post and Comment}: To post/comment on another user's wall, a user signs the reference to the post or comment with the append-policy signature key of the parent object, which she ABDecrypts from the parent object.
The average time to post or comment 
is $3.94$\,s 
The results are reasonable since a user can continue her OSN activities while the update is performed.

\subsection{Experiments on PlanetLab}
\begin{figure}[htbp]
\centering
\subfloat[View My Wall] {\includegraphics[width=0.62\columnwidth]{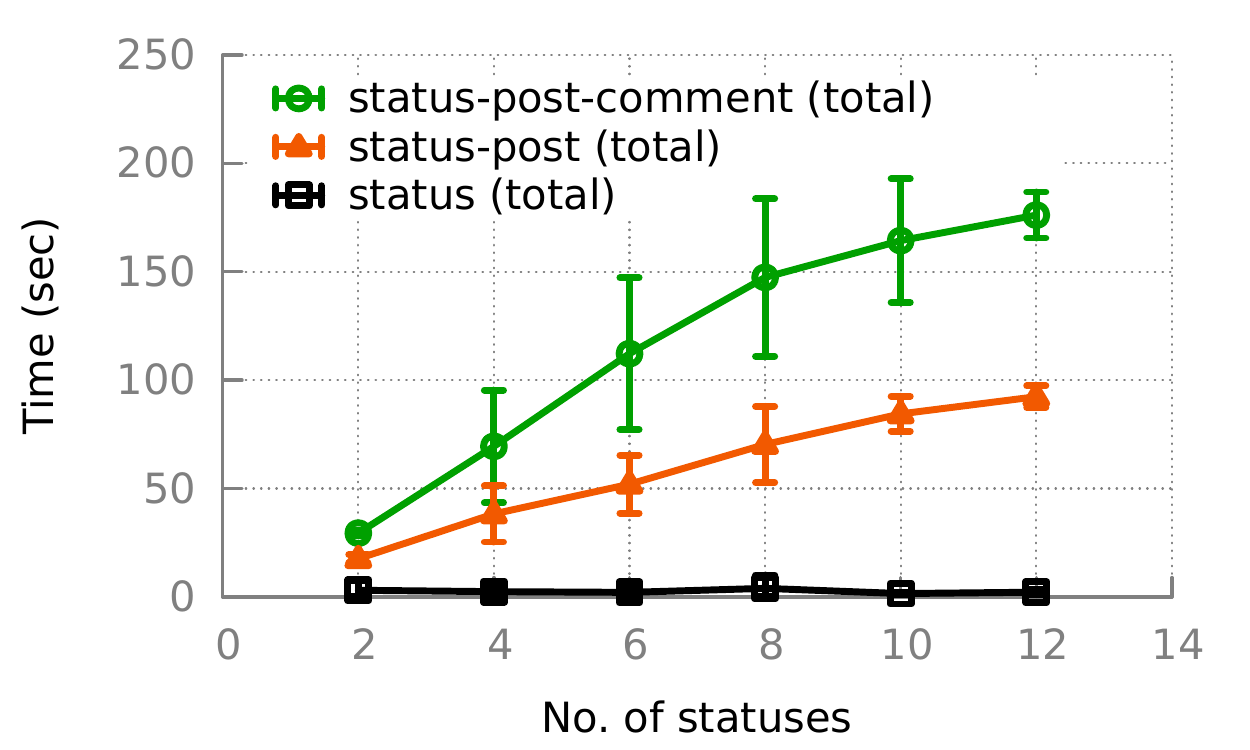} \label{chart:view-my-wall-pl} }\\
\subfloat[View Others' Wall] {\includegraphics[width=0.62\columnwidth]{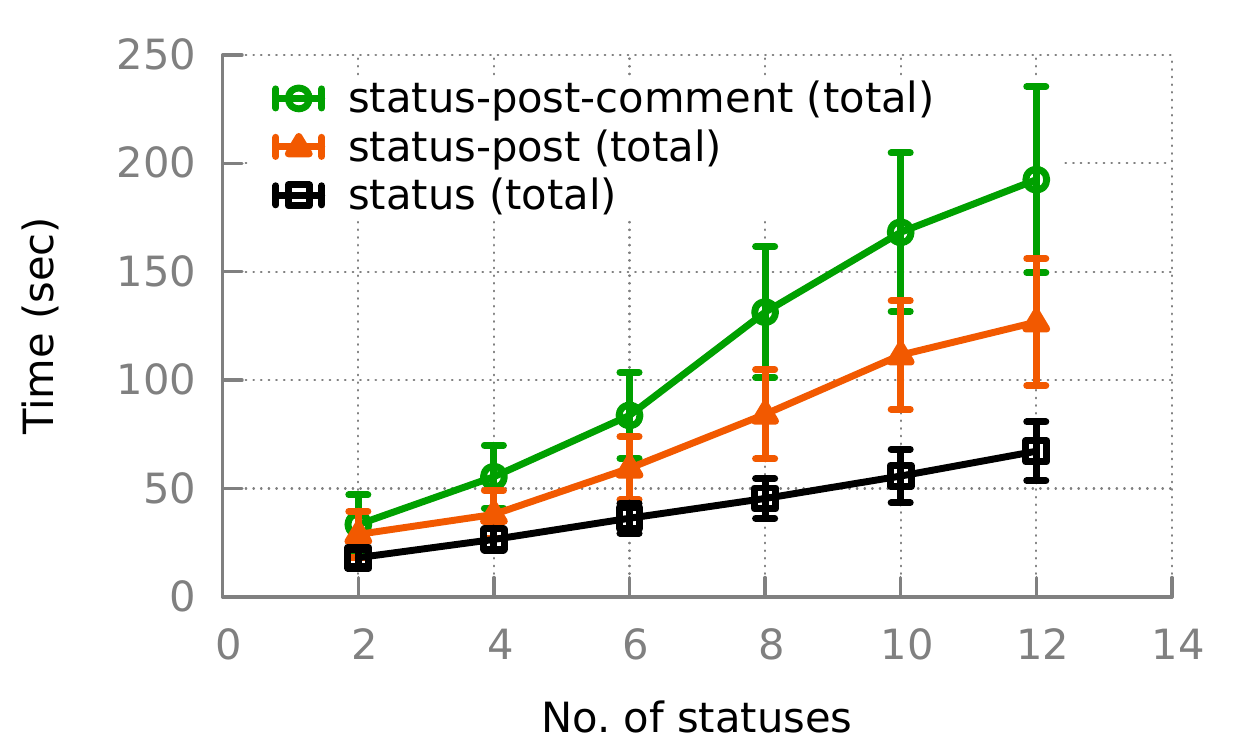} \label{chart:view-others-wall-pl} }
\vspace{-2mm}
\caption{PlanetLab Results: The average time to view others' wall with 10 statuses, posts, and comments is about 168~seconds. 
}
\label{chart:kad-pl}
\vspace{-5mm}
\end{figure}

We perform the same experiments on $15$ PlanetLab nodes to get an idea of \dosn's performance in a real deployment. 
Currently, the DHT has been implemented on PlanetLab using a Kademlia prototype extracted from the Likir implementation~\cite{Likir}.
Figure~\ref{chart:kad-pl} shows the results of our preliminary PlanetLab experiments. 
As expected, the time to view walls in PlanetLab machines takes slightly longer because of network delays, such as the communication between peers and the proxy.
In addition, because of node failures, a few of the users' walls could not be viewed, and in some experiments, walls were retrieved only partially. 
We also test the time to construct and view the newsfeed. A newsfeed with $11$ feeds takes $37.3$\,sec ($95\%$ confidence interval is $[34.4,40.1]$\,s),
 which closely resembles our simulation results. %
For improved performance and resilience, we are investigating the use of caching and replication parameters, which will be reported in subsequent work.

\section{Related Work}\label{sec:relatedwork}

Several projects such as Diaspora~\cite{diaspora}, PeerSon~\cite{buchegger09}, Safebook~\cite{Cutillo09}, and LotusNet~\cite{lotusnet} have addressed privacy in OSNs either through cryptography, architectural modifications, or decentralization of the provider. %
Diaspora is a social network that users install on their own personal web servers, without support for encryption. Backes et al.~\cite{Backes11}
present a core API for social networking, which can also constitute a plug-in for distributed OSNs. However, they assume that the server is trusted with the data while implementing access control protection. PeerSon, LotusNet and Safebook benefit from DHTs in their architecture. PeerSon and Safebook suggest access control through encryption, but they fall short in providing fine-grained policies comparing to ABE-based access control. Moreover, in all of these schemes overhead of key revocation affects performance. Safebook is based on a peer-to-peer overlay network named ``Matryoshka''. The end-to-end privacy in Matryoshka is provided by leveraging existing hop-by-hop trust. In LotusNet, which is based on Likir~\cite{aiello08}, the authors consider the distributed storage to be trusted and do not perform encryption. Likir uses signed grants to specify permissions and provide access control.

The closest work to \dosn is Persona~\cite{Baden09} that combines ABE with a decentralized OSN architecture to ensure data confidentiality. However, Persona falls short while supporting the fine-grained policy required for OSNs, because the underlying cryptographic mechanism~\cite{cpabe} lacks efficient revocation. \dosn uses an extended version of EASiER~\cite{easier} with support for access delegation while providing extremely fine-grained access control. In addition, \dosn provides user-verifiable data integrity through Attribute-based Signatures (ABS)~\cite{abs}. The storage service in Persona authenticates write operations through the requester's public key and hence can  learn the user's social contacts. \dosn solves this issue through carefully designed cryptographic techniques. Besides, unlike Persona, \dosn separates the policy used to encrypt the data from the ciphertext itself, thus preventing the  storage nodes or a third party from inferring a user's privacy policies by getting access to a ciphertext. Persona uses a multi-reader-writer service for the wall, but lacks a protocol for commenting on wall posts. \dosn provides a full design and implementation to read, write, and comment on walls or any data that appears on a wall. 

\section{Conclusion and Future Work}\label{sec:conclusion}

In this paper we proposed \dosn, a design for decentralized social networks with an emphasis on security and privacy.  \dosn uses an efficient cryptographic mechanism for confidentiality, combining traditional and advanced cryptographic schemes for integrity, and the use of a DHT for availability. We discussed the architecture in detail, and presented a prototype of our design. Simulation and experiments on PlanetLab with our preliminary prototype show that a privacy-enhanced OSN based on DHT with focus on confidentiality and integrity is a feasible architecture. Our future work includes adding more features to \dosn and further improving performance and resilience through optimized cryptographic techniques, caching, and replication.

\section*{Acknowledgments}
This research was funded in part by a grant from the Indiana University Center for Applied Cybersecurity Research.
\bibliographystyle{abbrv}
\bibliography{paper}    

\begin{thebibliography}{10}

\bibitem{diaspora}
Diaspora*.
\newblock \url{https://joindiaspora.com/}.

\bibitem{osn-selling-data}
Facebook, myspace confront privacy loophole.
\newblock
  \url{http://online.wsj.com/article/SB100014240527487045131045752567012154655%
96.html}.

\bibitem{eff-facebook-timeline}
Facebook's eroding privacy policy: A timeline.
\newblock \url{https://www.eff.org/deeplinks/2010/04/facebook-timeline}.

\bibitem{google-buzz}
Google seeks to quell buzz privacy outcry.
\newblock \url{http://benton.org/node/32202}.

\bibitem{nist}
{NIST}.
\newblock \url{http://csrc.nist.gov/groups/ST/toolkit/key_management.html}.

\bibitem{aiello08}
L.~M. Aiello, M.~Milanesio, G.~Ruffo, and R.~Schifanella.
\newblock {Tempering Kademlia with a robust identity based system}.
\newblock In {\em P2P}, 2008.

\bibitem{lotusnet}
L.~M. Aiello and G.~Ruffo.
\newblock Lotus{N}et: tunable privacy for distributed online social network
  services.
\newblock {\em Computer Communications}.
\newblock (to appear).

\bibitem{app:facebook}
{Facebook in Privacy Breach}.
\newblock
  \url{http://online.wsj.com/article/SB100014240527023047728045755584840752369%
68.html}.

\bibitem{Backes11}
M.~Backes, M.~Maffei, and K.~Pecina.
\newblock A security {API} for distributed social networks.
\newblock In {\em NDSS}, 2011.

\bibitem{Baden09}
R.~Baden, A.~Bender, N.~Spring, B.~Bhattacharjee, and D.~Starin.
\newblock Persona: an online social network with user-defined privacy.
\newblock In {\em ACM SIGCOMM}, 2009.

\bibitem{cpabe}
J.~Bethencourt, A.~Sahai, and B.~Waters.
\newblock Ciphertext-policy attribute-based encryption.
\newblock In {\em IEEE Security \& Privacy}, 2007.

\bibitem{buchegger09}
S.~Buchegger, D.~Schi{\"{o}}berg, L.~H. Vu, and A.~Datta.
\newblock {PeerSoN}: {P2P} social networking --- early experiences and
  insights.
\newblock In {\em SNS}, 2009.

\bibitem{castro}
M.~Castro, P.~Druschel, A.~Ganesh, A.~Rowstron, and D.~Wallach.
\newblock Secure routing for structured peer-to-peer overlay networks.
\newblock In {\em OSDI}, 2002.

\bibitem{flickr}
M.~Cha, A.~Mislove, B.~Adams, and K.~P. Gummadi.
\newblock Characterizing social cascades in {Flickr}.
\newblock In {\em WOSN}. ACM, 2008.

\bibitem{Likir}
U.~o.~T. Computer Science~Department.
\newblock {Likir}, 2008.

\bibitem{Cutillo09}
L.~A. Cutillo, R.~Molva, and T.~Strufe.
\newblock Safebook: Feasibility of transitive cooperation for privacy on a
  decentralized social network.
\newblock In {\em WOWMOM}, 2009.

\bibitem{green:usenix11}
M.~Green, S.~Hohenberger, and B.~Waters.
\newblock Outsourcing the decryption of {ABE} ciphertexts.
\newblock In {\em Usenix Security}, 2011.

\bibitem{gross}
R.~Gross and A.~Acquisti.
\newblock {Information Revelation and Privacy in Online Social Networks (The
  Facebook Case)}.
\newblock In {\em WPES}, 2005.

\bibitem{easier}
S.~Jahid, P.~Mittal, and N.~Borisov.
\newblock {EAS}i{ER}: Encryption-based access control in social networks with
  efficient revocation.
\newblock In {\em ASIACCS}, 2011.

\bibitem{whanau}
C.~Lesniewski-Laas and M.~F. Kaashoek.
\newblock Whanau: a {Sybil}-proof distributed hash table.
\newblock In {\em NSDI}, 2010.

\bibitem{abs}
H.~K. Maji, M.~Prabhakaran, and M.~Rosulek.
\newblock Attribute-based signatures.
\newblock Cryptology ePrint Archive, Report 2010/595, 2010.

\bibitem{Maymounkov02}
P.~Maymounkov and D.~Mazi\`{e}res.
\newblock Kademlia: A peer-to-peer information system based on the xor metric.
\newblock In {\em IPTPS}. Springer-Verlag, 2002.

\bibitem{Rowstron01pastry:scalable}
A.~Rowstron and P.~Druschel.
\newblock Pastry: {Scalable} distributed object location and routing for
  large-scale peer-to-peer systems.
\newblock In {\em Middleware}, 2001.

\bibitem{Singh09}
K.~Singh, S.~Bhola, and W.~Lee.
\newblock {xBook}: Redesigning privacy control in social networking platforms.
\newblock In {\em USENIX Security Symposium}, 2009.

\bibitem{tahoe}
Z.~Wilcox-O'Hearn and B.~Warner.
\newblock Tahoe: the least-authority filesystem.
\newblock In {\em StorageSS}, 2008.

\end{thebibliography}

\end{document}